\documentstyle[epsfig]{mnv2}
\def\gs{\mathrel{\raise0.35ex\hbox{$\scriptstyle >$}\kern-0.6em 
les
\lower0.40ex\hbox{{$\scriptstyle \sim$}}}}
\def\ls{\mathrel{\raise0.35ex\hbox{$\scriptstyle <$}\kern-0.6em 
ggles
\lower0.40ex\hbox{{$\scriptstyle \sim$}}}}
\def\ltorder{
\mathrel{\raise.3ex\hbox{$<$}\mkern-14mu\lower0.6ex\hbox{$\sim$}}
}
\def\gtorder{
\mathrel{\raise.3ex\hbox{$>$}\mkern-14mu\lower0.6ex\hbox{$\sim$}}
}
\def\msun{\>{\rm M_{\odot}}}
\def\deg{^\circ}

\title[Sunyaev-Zeldovich decrements with no clusters?]
{Sunyaev-Zeldovich decrements with no clusters?}
\author[Priyamvada Natarajan \& Steinn Sigurdsson]
{Priyamvada Natarajan \& Steinn Sigurdsson\\
Institute of Astronomy, Madingley Road, Cambridge CB3 0HA, U. K.\\}

\begin{document}
\label{firstpage}

\maketitle
  
\begin{abstract}
Sunyaev-Zeldovich decrements in the temperature of the Cosmic
Microwave Background (CMB) are produced by Thomson scattering 
of the CMB photons by the confined hot electrons in the Intra-Cluster 
Medium encountered along the line-of-sight. In this paper, we 
propose an alternate physical process that can produce detectable 
thermal and kinematic Sunyaev-Zeldovich decrements around quasars 
or quasar pairs in regions where no assembled, virialized clusters 
are detected either in the optical or X-ray wave-band. Invoking 
quasar outflows, we argue that both {\bf thermal} and {\bf kinematic} 
decrements can be produced by the baryons swept up by the mechanical 
luminosity of quasars. In contrast to the case of decrements produced 
by the hot electrons confined within the cluster potential, the 
magnitude of these quasar-outflow induced effects depend primarily 
on the luminosity of the QSO, and the physical scale of the outflow -
which in turn depends both on the mean local density and the density 
gradient in the vicinity of the quasar. Quasar outflows produce (i) 
a frequency independent kinematic decrement ${\Delta
T/T}\,\sim\,{10^{-4}}$; (ii) a frequency dependent fluctuation in the
intensity ${\Delta {J_{\nu}}/{J_{\nu}}}\,\sim\,$ a 20\% variation
relative to ${\Delta T/T}$, say at 15 GHz and 30 GHz; (iii) a temperature 
increment ${\Delta T/T}\,\sim\,{10^{-4}}$ on the opposite side 
of the QSO; (iv) a linear polarization signal over 20'' 
at $\sim\,{10^{-7}}$ level; and therefore (v) patches of 
signal ${\Delta T/T}\,\sim\,{10^{-4}}$ on degree scales which will 
contribute to the power and hence affect the CMB multipole 
calculations in the rising region of the first Doppler 
peak at $l\,\sim\,60$ for CDM-like models as well as defect models.
These effects are of interest since they are potentially observable 
in the context of the next-generation of CMB 
experiments - MAP \& PLANCK.
\end{abstract}
\begin{keywords}
cosmology: cosmic microwave background -- diffuse radiation --
galaxies: quasars -- general
\end{keywords} 
 
\section{Introduction}

Sunyaev-Zeldovich decrements in the cosmic microwave 
background are produced by the Thomson scattering 
of the photons by hot electrons - typically by electrons that are
confined in the dark matter potential well of a rich cluster of 
galaxies. However, we demonstrate in this work that comparable temperature 
decrements can be produced by electrons that are swept up by the 
outflows powered by the mechanical luminosity of quasars. 
While the properties of the outflow are determined primarily by the 
energy output of the QSO, the kinematics at late times are determined by 
the mean density and the density gradient in the vicinity 
of the QSO. Since in variants of the standard CDM (cold dark matter)
dominated cosmological models, quasars form from the collapse of 
high-significance peaks and tend to seed the high density  
regions which subsequently are the sites for the formation of clusters, 
the physical scale of the outflow depends on the ambient density and depth of
the local potential well. There are at present many new 
observations of the CMB in the millimeter wave-band planned or underway,
and there is renewed interest in the radio morphology 
of high-redshift QSOs and their hosts. In this paper, we expand on the 
classical S--Z scenarios (clusters, clusters with peculiar velocities, 
relativistic corrections to the scattering cross-sections and warm 
bubbles around QSOs with finite peculiar velocity) and include another
class of objects that may produce such decrements. For this
class of objects: quasar outflows, it is instructive to point out that 
the gas need not be very hot in order to produce 
an observable decrement. In particular, when the cooling/expansion 
time-scales are interestingly long, i. e. 
if the cooling time is comparable to or exceeds the time-scale of the high 
luminosity phase of the QSO, the effects of the gas can be observed well
after the QSO has faded. 

This paper is organized as follows: in Section 2 we motivate the role of 
quasar-driven outflows; in Section 3 the energetic 
constraints in terms of the implied physical 
scales out to which the outflows and their effects are likely to 
manifest themselves are outlined; in Section 4 we discuss the 
kinematic and the thermal S--Z decrement induced by outflows 
and their other observational implications; and we present our conclusions 
and prospects for detection in the final section.

\section{Quasar-driven outflows}

The ubiquity and high luminosity of quasars out to high redshifts
suggests that they might have played an important role in the 
subsequent formation of structure (Efstathiou \& Rees 1988; Rees 1993; Silk 
\& Rees 1998). Babul \& White (1991) suggest that the ionization produced 
by UV photons from the quasar might inhibit galaxy formation in their
vicinity due to the `proximity' effect. Thus quasar activity is likely
to be one of the important non-gravitational processes to have
influenced structure formation at redshifts $z\,>\,2$.

We propose quasar-driven outflows as a possible physical mechanism for 
concentrating baryons. Such a scenario has been explored previously
in the context of explosion induced structure formation models
(Ostriker \& Cowie 1981; Daly 1987) as well as in theoretical models of the
Inter-Stellar Medium (McKee \& Ostriker 1977). Mechanically powered
outflows can push baryons around active quasars in shells cleaning out 
regions of $R_{today} \gtorder 1 Mpc$. We examine the effectiveness 
of this mechanism in the production of massive clumps of baryons that can in 
turn produce a signal in the CMB. The
observational evidence for these outflows are seen in
a subset of radio-quiet QSOs - the broad absorption line systems 
(Hazard et. al 1984; Weymann et. al 1991) in which speeds v/c $\sim$
(0.01 -- 0.1) are inferred during a wind-phase. These `hyper-active' 
quasars can create both regions that are baryon-rich as well as voids 
via their energetic outflows. This powered outflow model
is compatible with both scenarios of quasar population evolution: the 
single generation of long-lived quasars as well as the multiple generation 
model wherein populations fade and flare (Haehnelt \& Rees 1994).
Superwinds of the kind detected by Heckman et al. (1996) from galaxies 
can also in principle produce S--Z decrements, but they are typically
an order of magnitude weaker than the QSO outflows proposed here. 

\section{Energetics of the Mechanism}

The total energy available to power a quasar can be directly estimated from the 
mass accreted by a black hole of mass $M_{\rm bh}$ (modulo an efficiency 
factor, $\epsilon$). In general, the total energy available is 
10-20\% of the rest mass of the central black hole ($\epsilon \sim 0.1 - 0.2$).
We assume that 
roughly 50\% ($\epsilon_L \sim 0.5$) of the total energy in turn might 
be available as mechanical power. 
By assumption this sets up an outflow that is primarily mechanically powered,
that is, a fraction of the
energy emitted by the active quasar gets efficiently converted into kinetic
energy of the intergalactic medium.
Mechanical luminosities up to
$\sim 10^{48}\, {\rm erg\, s^{-1}}$ at redshifts $z \gtorder 4$ for
such QSOs are postulated, this is not entirely implausible given 
inferred X--ray luminosities of comparable magnitude for high $z$ 
QSOs (Elvis et. al 1994). Clearly such energy coupled locally to the
environment as mechanical luminosity would set up an 
outflow that sweeps out a region of scale $R$:
\begin{eqnarray}
R\,\sim\,{1}\,({\frac {{\epsilon}\,\epsilon_L}{0.05}})^{\frac 1 3}\,({\frac {M_{\rm
bh}}{10^{11}\,\msun}})^{\frac 1 3}\,({\frac {f_Q}{1}})^{-\frac 1
3}\,({\frac {v/c}{0.01}})^{-\frac 2 3}\, Mpc.
\end{eqnarray}
For our purposes typical swept up regions are of the order of 0.1--1 Mpc or so; 
$f_Q$ is the local baryon fraction in units of 0.01, and a
density contrast of $\sim 10^3$ over the mean cosmological density
at redshift $\sim 3$ is required for the numbers to be consistent, this
is a plausible over-density for the highest significance peaks that will seed 
the massive clusters observed at lower redshifts.
The total energy budget
${E_T}\,\sim\,{{\epsilon\,{\epsilon_L}}{M_{\rm bh}}}$; and the 
restrictions on the available parameter space in terms of the mass 
that needs to be swept up and the bulk velocity required to produce 
a kinematic S--Z decrement are dictated by the following relations,
\begin{eqnarray}
{{E_T}\,\sim\,{{M_{\rm sweep}}{{v_{\rm
sweep}}^2}}}\,;\,\,{{R}\,\sim\,{{t_{\rm QSO}}\,{v_{\rm
sweep}}}}\,;\,\\\,{{\frac {\Delta T}{T}}}\,\sim\,{{\tau}{\frac
{v_{\rm sweep}}{c}}}\,\,\sim\,{M_{\rm sweep}}\,{v_{\rm sweep}}\,,
\end{eqnarray}
where $t_{\rm QSO}$ is the quasar lifetime and $\tau$ the optical
depth.
Beam dilution imposes a minimum $M_{\rm sweep}$ assuming the line
of sight length through the gas is comparable to its transverse
scale, which is likely for the resulting outflow geometry studied here.
Scaling to typical Abell cluster parameters, the
mean electron thermal 
speeds are $\beta_T = v_T/c \sim 0.1$.
Since the kinematic effect is first order in $\beta_K = v/c$,
we require $\beta_K \approx \beta_T^2 \sim 10^{-2}$. This is consistent with
speeds required for the mass of swept up gas which would produce the
required optical depths over 
angular scales of a few arc minutes. It is these mass and velocity 
requirements which
drive us to postulate ultra--massive central black holes to power the
outflows.
Note that ${{\tau}\,\sim\,{{n_e}R}}$ whereas the mass 
swept up ${{M_{\rm sweep}}\,\sim\,{{n_e}{R^3}}}$. All the numbers 
quoted in this work have been computed for
${H_0}\,=\,50\,{km\,s^{-1}\,Mpc}$; ${\Omega_0}\,=\,1.0$ and $\Lambda\,=\,0$.
As illustrated by eqns. 2 and 3, the best sources for large 
energy effects are the massive BHs that power the QSOs.
We also note here that according to some models of unified 
schemes for AGNs, jets and efficient energy release occurs only if the 
QSO is accreting close to or at super-Eddington rates, possibly via
the Blandford-Znajek mechanism for rotating BHs. This might 
preferentially happen at
high redshift, where a higher fraction of high luminosity QSOs are observed,
consistent with a picture wherein the most massive BHs form rapidly at high
redshift and are fueled by the baryons that are present in these high-density
regions. We do not necessarily need fully assembled, virialized dark halos 
for this to happen, we only need an efficient way to funnel the baryons already
there to the
bottom of the deepest potential wells. 

\subsection{Dynamics of the quasar outflow}

The outflow starts out being collimated close to the
source and has some, small, opening angle so that it does
not terminate or disrupt the accretion process which is its
powerhouse. At a distance of $\gtorder\,10\,kpc$ away from the central
engine, the flow opens out. The mass $M_{\rm sweep}$ (of the IGM)
swept out by the outflow is: 
\begin{eqnarray}
M_{\rm sweep}\,\sim\,{L_{\rm QSO}};\,\,\,\,v_{\rm sweep}\,\sim\,{\sqrt({L_{\rm QSO}})},
\end{eqnarray}
where ${L_{\rm QSO}}$ denotes the mechanical luminosity of the QSO.
As was shown by Phinney (1983), jets from QSOs start out mildly 
relativistic with a Lorentz factor $\sim$ a few; it is likely that the
initial outflow is dominated by a pair plasma. The total momentum
that the outflow can deposit is determined primarily by the ratio 
of ${\frac {m_p}{m_e}}$ and therefore bulk speeds of the order 
of $v\,\,\sim\,{10^{-2}}c$ are expected, which are indeed observed
at low redshifts even in low power radio galaxies like NGC 315 (Bicknell 1994). 
The mass swept out by a
single QSO with an outflow assuming that 50\% of its total bolometric 
luminosity is expelled mechanically is,
\begin{eqnarray}
{M_{\rm sweep}}\,&\sim&\,\nonumber{2\,\times\,{10^{14}}}\,({\frac
{L_{QSO}}{10^{48}\,erg\,s^{-1}}})\,({\frac
{\delta t}{{5\,\times\,10^8}\,yr}})\,\\&\times&({\frac
{v_{sweep}}{3000\,km\,s^{-1}}})^{-2}\,\msun,
\end{eqnarray}
where $\delta t$ is the duration of the outflow phase, $v_{\rm sweep}$
the ejection velocity and $L_{QSO}$ the mechanical
luminosity of the QSO. The total mass swept up is strictly bounded,
\begin{eqnarray}
{M_{\rm sweep}}\,\leq\,{\frac {2{\epsilon}\,{\epsilon_L}\,{M_{\rm bh}}}{(\frac {v_{\rm sweep}}{c})^2}}.
\end{eqnarray}
While the initial outflow is likely to 
be a pair plasma, the number density of electrons in the bulk 
outflow is dominated by the swept up hydrogen from the IGM and is 
estimated as,
\begin{eqnarray}
{n_e}\,\sim\nonumber\,1\,\times\,{10^{-2}}\,\biggl({\frac
{M_{\rm  sweep}}{2\times 10^{14}\,\msun}}\biggr)\,({\frac
{R }{1\,Mpc}})^{-3}\,{cm^{-2}},
\end{eqnarray}
in good agreement with typical values obtained using 
the 151-MHz and 1.5-GHz data set of Leahy, Muxlow \& Stephens (1989). 
The optical depth $\tau$ to Thomson scattering is:
\begin{eqnarray}
\tau\,\sim\,{10^{-2}}\,\,
({\frac {n_e}{10^{-2}\,cm^{-3}}})\,\,({\frac {R}{1\,Mpc}}).
\end{eqnarray}
With an approximate slab geometry as predicted in this outflow picture the gas 
mass required for a given optical depth can be several times
smaller than that for an isothermal sphere whose virialized core
provides the most of the optical depth for the clusters S--Z effect.
No dark matter is swept up, so the total mass involved is
considerably less than the total cluster mass of a virialised
cluster producing a comparable S--Z effect.
As is evident from our
approach, the micro-physics of the early stages of the jet propagation 
have not been computed in detail. We assume that the flow is initially 
collimated and breaks out on a scale of a few kpc and that 
approximately 50\% of the energy released due to accretion is efficiently 
mechanically coupled to the environment. The geometry of a typical region
through which the outflow propagates is assumed to be a 
cosmologically overdense region with a strong external density 
gradient. Physically, this is required so that the initial outflow can 
remain collimated right out to the advancing surface where 
the bulk of the energy is deposited. The advancing surface breaks
out and expands once a critical ambient pressure is attained 
and the presence of the density gradient serves to anisotropize (i. e. 
destroys the spherically symmetry of) the post break-out expansion.   

\section{S--Z decrement produced by the outflow}

The S--Z decrement in the temperature of the CMB is produced by
Thomson scattering of the CMB photons by hot electrons encountered 
along the line-of-sight (Sunyaev \& Zeldovich 1970; 1972; 1980). The
magnitude of the decrement is independent of the redshift of the hot
plasma and is given by the integral of the pressure along the
line-of-sight. The effect has two components a {\bf{thermal}} and a
{\bf{kinematic}} one (Sunyaev \& Zeldovich 1980). The thermal S--Z
effect causes fluctuations in the temperature and the intensity which
are (i) frequency independent, of equal magnitude and the same sign in
the Rayleigh-Jeans limit of the spectrum but (ii) are larger, differ
in magnitude and depend on the frequency in the Wien limit. The
overall sign of the contribution to the fluctuations (in both the
temperature and the intensity) also differs in the Rayleigh-Jeans and
the Wien limit. 

\subsection{The Kinematic S--Z effect}

Coherent streaming motion with a velocity $v_{\rm
sweep}$ with respect to the frame of reference in which the CMB is
isotropic introduces a Doppler shift in the scattered radiation
producing the {\bf{kinematic effect}} which additionally perturbs the
temperature and intensity across the region. This kinematic piece
induces a frequency {\bf{independent}} temperature gradient but a
frequency {\bf{dependent}} fluctuation in the intensity; both of which
have the same sign (determined by the direction of the velocity
vector; note that if the outflow is spherical on large scales the
kinematic S--Z effect cancels as the line of sight passes through
regions with opposite sign radial velocity).  Sunyaev \& Zeldovich
(1980); Peebles (1993), show that the decrement produced by the kinematic
term is given by,
\begin{eqnarray}
{\frac {\Delta T}{T}}\,\approx\,-{\tau}\,({\frac {v_{\rm sweep}}{c}})\,\,;
{\frac {\Delta {J_{\nu}}}{J_{\nu}}}\,=\,{\frac {\Delta
T}{T}}\,{\frac{x\,exp(x)}{exp(x) - 1}};
\end{eqnarray}
where $x={\frac {h \nu}{k T_{\rm r}}}$; the kinematic decrement is first order 
in ${v_{\rm sweep}}/c$ compared to the normal second order thermal
effect produced by hot cluster gas. For typical values of the
optical depth and the velocity of the outflow assumed in the above
analysis, this yields a 
${\frac {\Delta T}{T}}\,\sim\,{3\,\times{10^{-4}}}$. We stress here that 
{\bf{the ratio of the predicted thermal to the kinematic S--Z
effects depends entirely on the bulk velocity and mean temperature
of the plasma}}. Below we
estimate the thermal component of the S--Z effect expected from a fiducial 
outflow. 

\begin{figure}
\centerline{
\psfig{figure=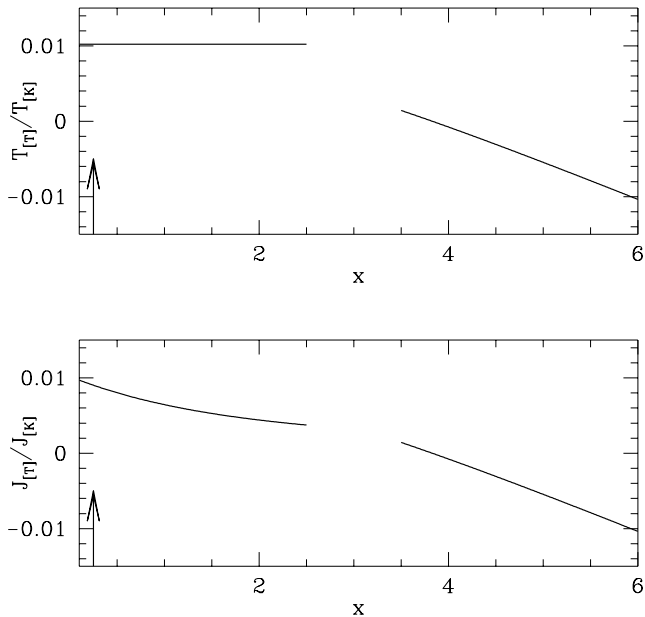,width=0.5\textwidth}
}
\caption{The ratio of the predicted thermal decrement (denoted by the
subscript [T]) to the kinematic component (denoted by the subscript
[K]) of the decrement in the temperature and intensity in the
Rayleigh-Jeans limit [left side of the graph] and the Wien limit [right side] 
plotted as a function of the dimensionless parameter x, defined as
$x\,=\,{\frac{h\nu}{kT_r}}$ (where $T_r$ is temperature of the
radiation i. e. 2.7 $\,K$). The ratios plotted above have been
computed for the plasma at $T_e$ = 10$^6\, K$ and an optical depth
$\tau$ = 0.01. The arrow on the x-axis marks the value corresponding 
to a frequency of 15 GHz. The x range varies from roughly $\lambda$
= 5 cm at x = 0.1 to $\lambda$ = 1 mm at x = 5.0.}
\end{figure}
The material at the working surface may be shock-heated to temperatures
as high as $\sim\,{10^8}\,{K}$, but may cool down efficiently as 
we argue below. The bulk of the mass (which is the material in the
IGM; note that only roughly a thousandth of the mass is expelled directly from 
the QSO) however is expected to be in a cooler phase. The shock
heated dense plasma will cool to a temperature $T_e$ such that its cooling
time $t_{\rm cool}\,\sim\,{t_{\rm dyn}}$, where ${t_{\rm dyn}}$ is its
dynamical time (Rees \& Ostriker 1977). Thermal brehmstrahlung is expected 
to be the primary process for energy loss in the early stages of the 
evolution of the bubble and X-ray fluxes $\leq 10^{44} erg s^{-1}$ are 
expected, which could be detectable for  
low-redshift bubbles (see Natarajan, Sigurdsson \& Silk 1998
for a more detailed discussion). X--ray cooling is inefficient for
these high power outflows.
The plasma can cool to lower
temperatures rapidly via turbulent mixing which will tend to homogenize the 
temperature of the material on the dynamical time-scale. Synchrotron cooling
is however, unlikely to be important, since for the typical estimated value of 
the IGM magnetic field at high-redshifts $|B_{\rm IGM}| \sim {10^{-8.5}}$ G  
(see Kronberg 1994) the cooling time exceeds both the Hubble time and 
the lifetime of the QSO, therefore, unless there are significantly stronger 
B fields on very small scales energy losses via synchrotron processes will 
be negligible. The thermal S--Z decrement produced by this warm plasma is:
\begin{eqnarray}
{\frac {\Delta T}{T}}\,=\,({\frac {2\,k\,T_e} {m_e\,c^2}}\,\tau)\,
=\,{3.45\,\times\,{10^{-6}}}\,({\frac {T_e}{10^6\, K}}),
\end{eqnarray}
much lower than that expected for virialized gas in 
a core of a cluster of galaxies. For a range of frequencies, in Fig. 1, 
we plot the ratio of the expected kinematic and thermal components for 
the plasma in the outflow. As shown by Birkinshaw
(1996), the ratio of the thermal to the kinematic effect is independent 
of the optical depth and can be written as,
\begin{eqnarray}
{\frac {{\Delta T}_{K}}{{\Delta T}_{T}}} \,=\,{40}\,({\frac {v_{\rm
sweep}}{3000\,km\,s^{-1}}})\,{}\,({\frac {T_e}{10^6\,K}})^{-1},
\end{eqnarray}
therefore, the contribution from the kinematic effect could easily be
one or even two orders of magnitude higher than the thermal effect,
for warm ($T \sim 10^6\, K$)
plasma with significant bulk motion, as postulated here.

This mechanism might be relevant to a 
recent detection of a decrement in the region of the quasar 
pair PC1643+4631 where no cluster is detected. The reported 
observation was obtained by Jones et. al (1997) using the 
Ryle telescope at 15-GHz in the direction of the quasar pair 
PC1643+4631 A\& B. The pair separation on
the sky is $\sim$ 200'' and their $r_4$ magnitudes are 20.3 and 20.6
respectively.
The redshifts of the quasars are (A) $z= 3.79$ and (B) $z=3.83$ 
(Schneider, Schmidt \& Gunn 1994).
The central value of the measured decrement is 560
$\mu$K. Interpreting the source as a cluster at $\sim$ 5 keV, Jones
et. al estimate the implied gas mass to be $\gtorder$ 2 $\times$ 10$^{14}$
$\msun$ and conclude that the result demonstrates the existence of a
massive high redshift (z $\gtorder$ 1) system. However, serendipitous and
pointed ROSAT PSPC observations (Kneissl et. al 1997) as well as 
ground-based optical follow-ups (Saunders et. al 1997) have failed 
to reveal a cluster of galaxies as the source of the hot plasma to 
produce this `cold' spot in the CMBR.
The hypothesis of lensing by an intervening massive 
cluster is therefore hard to justify. Another S-Z decrement has 
been measured at the VLA (3.6 cm) by Richards et. al 1996, also 
around a pair of radio-quiet QSOs at a projected separation of 1 Mpc. 
In this case the pair have the same redshift (z = 2.561) and although 
their spectral features are not identical their velocity separation 
being small, they are likely to be a lensed pair split by a high-redshift 
cluster with a large time delay of the order of 500 years. 
The physical scales involved in both these observations are 
consistent with the role of quasar outflows and therefore our proposed model 
might provide a possible physical explanation.

\subsection{The Thermal S--Z effect}

\subsubsection{Emission from the shock-heated bubble}

In a recent analysis Aghanim et. al (1996), consider another scenario  
where the kinematic S--Z effect is also much higher than the
thermal S--Z effect. They consider the bulk motion of highly ionized baryons
in the proximity of QSOs at speeds of
$\sim\,{300\,km\,s^{-1}}$ and at temperatures of $\sim\,{10^5}\,{
K}$, essentially stationary bubbles around the QSO moving 
with the QSO at the local peculiar
velocity. Although the physical picture that we present above is entirely
distinct from that proposed by Aghanim et. al, we find the same
scaling of the decrement with the relevant parameters:
\begin{eqnarray}
{\frac {\Delta T}{T}}\,\sim\nonumber\,{3\,\times\,{10^{-4}}}\,
({\frac {\tau}{10^{-2}}})\,({\frac
{v_{\rm sweep}}{3000\,km\,s^{-1}}})\,({\frac {L_{\rm QSO}}{10^{48}\,erg\,s^{-1}}})^{\frac 1 3}\,({\frac {1 + z}{1 + 3}})^2.
\end{eqnarray}
The magnitude of the kinematic S--Z effect from outflows scales 
primarily with the mechanical luminosity of the QSO and the redshift,
therefore, we expect stronger decrements preferentially from high 
luminosity and high
redshift QSOs. Bicknell (1994) finds for a local FRI radio galaxy 
NGC 315, 
$L\,\sim\,{10^{43}}\,erg\,s^{-1}$; 
$\beta\,\sim\,{10^{-2}}$ and an associated
jet that extends to scales of the order of $R\,\sim\,{0.15}\,Mpc$; our
model considers QSOs that are essentially the high-redshift,
scaled-up (scaled up in luminosity and hence the physical scale on 
which the outflow occurs) versions of objects like NGC~315. 

We stress here that
spherically symmetric outflows do not generate a kinematic S--Z signal, 
an anisotropy induced by the bubble expansion 
into a density gradient is required.
More importantly with every observed kinematic
S--Z decrement a corresponding increment is produced due to the opposite  
sign bulk velocity on the other side of the QSO.

Several immediate predictions follow from our proposed scenario: while
the temperature decrement is 
frequency independent, the intensity change is not (shown explicitly in
eqn. 8 above), so we expect a $\sim 20\%$ variation 
in $\Delta J/J$ relative to $\Delta T/T$ at say 15 GHz and 30 GHz. 
We also expect a 
temperature $\bf increment$ of the same amplitude on the opposite 
side from the  decrement (see the marked `hot' spot HS and the cold spot as 
CS in Fig. 2), as we expect an outflow with the opposite sign radial velocity
on the other side of the QSO.

\begin{figure}
\centerline{
\psfig{figure=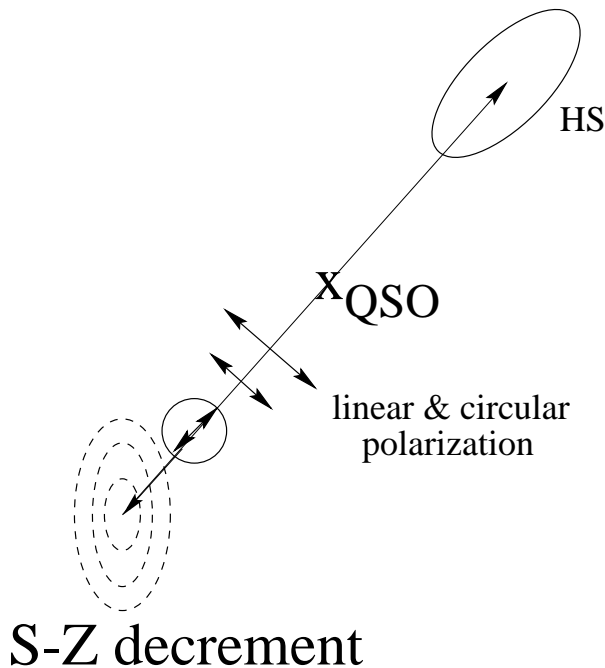,width=0.5\textwidth,angle=270}
}
\caption{Schematic of the configuration showing the relative positions
on the sky of the QSO; the observed S--Z decrement CS; the predicted
temperature increment HS.} 
\vspace{-0.5cm}
\end{figure}

Since the optical depth $\tau$ and $v/c$ are 
of comparable order, both polarization effects (see Sunyaev \&
Zeldovich 1980) arising due to (i) the
finite angle subtended by the line-of-sight with respect to the
velocity vector (linear polarization - $\sim\,{0.1\,{\tau}{(\frac
{v}{c})^2}}$) and (ii) the finite optical depth (circular
polarization $\sim\,\pm{0.1\,{\tau^2}{\frac {v}{c}}}$) are also roughly
equal. Therefore, we expect polarization at the $10^{-7}$ level on
20'' scales from flow transverse to the line of sight, along the line
from the QSO to the peak decrement. Given the density
and the temperature of the outflow material, the corresponding X-ray
luminosity $L_X$ is expected to be $\ltorder\,{10^{44}\,erg\,s^{-1}}$, 
below the flux limit of current X-ray observations, and substantially
redshifted to lower energies.
We also note that the presence of extended cool gas with 
$n_e\,\sim\,{10^{-2}}$ as predicted by this outflow picture could 
be detected via absorption features seen blueward of the emission
peak (offset by $\sim 3000$ km s$^{-1}$) in the QSO spectrum. 
Given the available energy budget,
diffuse Lyman-$\alpha$ emission from recombination in the warm gas
is expected with approximate surface brightness
within a narrow band filter 
(width $\sim$ 60 Angstrom)
that is of the order of $\sim\,{25}\,mag\,arcsec^{-2}$. 
Such diffuse emission may be detectable through
narrow band imaging or long slit spectroscopy, it is interesting to
note that somewhat smaller scale diffuse emission has been seen around
both a low redshift, low luminosity radio--loud QSO 4C 03.24 
(van Ojikpine
 et al. 1997) as well as in two high redshift radio--loud 
QSOs (Lehnert \& Becker 1998). The diffuse emission predicted for 
the QSO pair (PC1643+4631) is of comparable power to that seen 
around the quasar 4C 03.24, but spread over a larger 
angular scale due to the greater spatial extent required of 
the outflow. Any fortuitous alignment of two QSOs such that their 
outflows contribute additively to the S--Z decrement would make 
the effect unusually strong and hence detectable. We would expect this 
effect to be rare in general, even in the proximity of bright, high 
redshift, radio-quiet QSOs with the optimal opening angle 
($\theta\,\approx\,45\deg$). It also strongly depends on the 
efficiency of the conversion of the bolometric QSO 
luminosity to mechanical outflow with time and environment. The
luminosity of radio quiet QSOs is seen to increase 
strongly with $z$ (Barvainis, Lonsdale \& Antonucci 1996; Taylor 
et. al 1996).
For instance, in the scheme of Falcke et. al (1996) the
QSOs in PC1643+4631 at their brightest would have been classified
as FRIIs with luminosities 
in excess of $10^{48}\,erg\,s^{-1}$ (see Falcke et. al's Figure 1); 
the radio jets have now faded and the accretion power decreased by
many orders of magnitude, possibly
due to depletion of the cooling flow in the inner kpc (see eg. Fabian 
and Crawford 1990) or decreased efficiency at lower mass accretion rates. 
Since the outflow must not be spherical at large
radii for a kinematic S--Z effect to be observable, substantial amounts
of gas ought to remain in the vicinity of the QSO in the regions outside the
out-flowing cone, and would probably condense into the lyman absorbers or 
even a population of baryon-rich dwarf galaxies (Natarajan, 
Sigurdsson \& Silk 1998) observed in the vicinity of QSOs.

The presence of bright radio sources in the immediate
vicinity could make this effect difficult to detect. In the context
of unified AGN models, the viewing angle implied for radio loud 
QSOs means looking straight down the jet, therefore detection of such 
extended bulk flows from them is impossible. Hence one would only
see such effects after the peak emission had diminished by several orders
of magnitude and the extended radio lobes faded. 

In order to predict how many such objects are expected to be detected, we 
need to estimate the number density of QSOs which had
strong outflows. If we believe the jet 
power scalings with luminosity and redshift
predicted by models, the relevant number density is then of   
QSOs with luminosities $\gtorder\,{10^{47}\, erg\,s^{-1}}$ within 
the optimum redshift range $z\,\geq\,1$. In order to not
violate (the admittedly weakly constrained) local BH mass functions,
we expect the ultramassive BHs 
needed for this scenario to have a mean space density of ${10^{-7}}$ per
Mpc$^{-3}$ or less.

\section{Conclusions}

S--Z decrements measured in the vicinity of QSOs where no clusters
are detected either in the optical or the X-ray  
may be due to a concentration of baryons aggregated by
the power of mechanical outflows from these quasars. The above
phenomenon leads to the following predictions (note
that the numbers below are for the {\bf{kinematic}} S--Z decrement 
caused by the outflow from a {\bf{single}} QSO):
\begin{itemize}
\item (i) it produces a frequency independent decrement ${\Delta
T/T}\,\sim\,{10^{-4}}$; 
\item (ii) a frequency dependent fluctuation in the
intensity ${\Delta {J_{\nu}}/{J_{\nu}}}\,\sim\,$ a 20\% variation
relative to ${\Delta T/T}$ at say 15 GHz and 30 GHz; 
\item (iii) a 
temperature increment
${\Delta T/T}\,\sim\,{10^{-4}}$ on the opposite side of the QSO; 
\item (iv)
a linear polarization signal over 20'' at $\sim\,{10^{-7}}$ level
(pushing at the limits of the detection by the proposed new instruments on MAP
and PLANCK); 
\item (v) patches of signal ${\Delta T/T}\,\sim\,{10^{-4}}$ on
degree scales which will contribute to the power and hence affect the 
CMB multipole calculations in the rising region of the first Doppler 
peak at $l\,\sim\,60$ for CDM-like models as well as defect models; 
\item (vi) we expect this to be a rare phenomenon,
occurring only in 1 -- 3 \% of
all high-redshift QSOs; and finally, 
\item (vii) the thermal effect is
one or two orders of magnitude smaller than the 
kinematic one. 
\end{itemize}
In principle, the effect ought to be observable for QSOs in 
cases with the appropriate opening angle and alignment of
outflow to the line of sight, and preferentially for 
high mechanical luminosity, high redshift, radio-quiet
QSOs. We note with interest the very large scale diffuse
radio emission seen on two sides of A3667 (R\"ottgering et al. 1997).
If the source there is due to outflow from a now quiescent black
hole in the central galaxy of the primary cluster, which is one of
the possible explanations, then future observations should reveal a
supermassive black hole in the cD, primarily through a sharply
rising central dispersion in the inner regions of the galaxy. From
the scale of the outflow, if it was powered by an AGN outflow,
a black hole mass of $\sim 10^{10} \msun$ accreting at near Eddington
luminosity for at least a Salpeter time is required, and the activity
would have peaked $\gtorder 10^9$ years ago.

Since the physical model that we have presented here is 
speculative, we briefly summarize below the problems presented 
by the observations of S--Z decrements in QSO fields with no detected 
clusters:

(i) A wind-driven model composed of a pair plasma is somewhat
analogous to the state of the bubble during the early phases of 
our proposition here but such material is too hot and would
have a short lifetime, and, more
importantly, the maximal scale out to which such plasma could be swept out 
is only of the order of $\sim\,$ 100 kpc, insufficient to explain decrements 
on large scales $\sim\,$ 1 Mpc. Furthermore, such plasma should 
produce easily observable diffuse radio emission as it collides 
with the IGM. The observational implications of this variant have been
discussed in Natarajan, Sigurdsson \& Silk (1998).

(ii) It is possible that decrements can be caused by warm gas in 
the potential well of proto-clusters. The gas would not be in 
virial equilibrium, such a scenario might fit with the 
essential outflow model we present here, but in the limit of a more 
modest spherical outflow. Any detected decrement would then be 
the signature of the hot gas formed by an asymmetric collision 
of the unconfined outflowing gas with cold infalling gas.
If the infall is strongly non--spherical, as is likely in unvirialized
proto--clusters, this would produce a hot spot,
therefore producing a one-sided thermal SZ effect.
Such a scenario is therefore easily distinguished from the pure kinematic
effect discussed above, as the thermal decrement would show as
a strong increment at shorter wavelengths and the predicted
long wavelength
increment on the opposite side of the QSO would be absent.

Although we have presented a somewhat speculative model, it 
makes several falsifiable predictions.
Our model does require ultra-massive black holes, but that is not forbidden,
in fact observations imply precisely that some of the high redshift
QSOs have total luminosities in excess of $10^{48} \, {\rm erg\, s^{-1}}$,
implying central black hole masses of the order of $10^{10}\, \msun$.
We note that the estimated BH mass in the case of the nearby BL Lac
OJ287 (Lehto \& Valtonen 1996) 
at z = 0.306 is $\sim\,2\,\times\,{10^{10}}\,\msun$, and this is 
unlikely to be the most massive black hole in the universe.
How ultramassive black holes could form at high redshift to produce
such high luminosity sources; and whether high luminosity QSOs are
particularly efficient at producing mechanical outflows are separate
issues to be contemplated if further observations are consistent
with the model proposed here.

\section*{ACKNOWLEDGEMENTS}

We thank Alastair Edge, Andy Fabian, Mike Jones, Richard McMahon,
Martin Rees and Richard Saunders for valuable 
discussions. SS acknowledges the support of the European Union, through a TMR 
Cat. 30 Marie Curie Personal Fellowship. We thank the referee Sterl
Phinney for invaluable feedback that led to significant 
reorganizing of the paper.

\end{document}